# Towards All Optical, Universal Quantum Computation using Trapped Electron Spins and Cavity Polariton Resonance

Shruti Puri[1], Na Young Kim[1], Eisuke Abe[1,2] and Yoshihisa Yamamoto[1,2]
*1. E. L. Ginzton Laboratory, Stanford University, Stanford, California 94305, USA and*
*2. National Institute of Informatics, 2-1-2 Hitotsubashi, Chiyoda-ku, Tokyo 101-8430, Japan*

We propose an all optical quantum computation scheme, with trapped electron spin qubits, using their Coulomb exchange interaction with optically excited microcavity exciton-polaritons. This paper describes a single qubit rotation, which together with two-qubit controlled-$z$ gate presented in PRB **85**, 241403(R) (2012), form a set of universal logic gates. The errors due to finite cavity lifetime and incorrect orientation of the rotation axis are minimized by optimizing pump pulse parameters. With projective homodyne phase measurement and initialization, our scheme is a promising candidate for the physical realization of a universal quantum computer.

DiVincenzo established the five requirements for any candidate for quantum computer implementation [1]. Physical realization of a quantum computer is a challenge because of the inherent conflict between these criteria, which demand a strong coupling to control and measure the quantum bits (qubits) and at the same time require a strong isolation from the uncontrolled environment to preserve coherence. There exist many proposals based on: photon qubits, which suffer from imperfect single photon source, detectors and photon loss; nuclear spins, which lack in initialization and measurement capabilities; ion trap and superconducting qubits, but it is uncertain if these could ever be scaled up to perform truly large scale quantum computation because of increased decoherence; NV centers in diamond, although it is difficult to fabricate optical cavities to link the NV centers [2]. With unprecedented growth in technology, electron spin qubits in semiconductor quantum dots (QDs) are emerging as promising candidates for implementation of a scalable quantum computer. The ultrafast optical control of a single spin in QD has been demonstrated [3]. The challenge however is to establish a two-qubit gate which requires interaction between two electron spins. One proposal for such an operation demands unrealistic value of the optical cavity quality factor Q [4].

We propose a new technique in which the localized electron spins interact with optically excited, quantum well (QW) exciton-polaritons, via spin dependent Coulomb exchange interaction. This interaction, resembling the electron-electron Ruderman-Kittel-Kasuya-Yosida (RKKY) coupling [5], has the advantage of being strong, because of its electronic nature and also long-ranged, since the polariton is $10^4$ times lighter than an exciton [6]. In addition, the light mass of the polaritons reduces the interaction with the environment such as phonons, making the polaritons an ideal interaction medium. [7].

A similar scheme using electron spins and microcavity exciton-polaritons has been proposed [8, 9], in which both bright and dark excitons are used to achieve quantum logic operations. In contrast, the absence of dark exciton states in our scheme precludes the decoherence caused by spin relaxation processes, leading to gate fidelities of up to 99.99% [10, 11]. Here, the major channel for loss of coherence is the finite polariton lifetime in the cavity. Significant decoherence in the system can leak the information about the state of the system to an eavesdropper. In addition to analyzing dependence of fidelity on various parameters of a practical setup, this paper explains how such a loss of *which-path information* is *erased*.

The system we propose consists of a single GaAs QW, bounded by AlGaAs layers, placed in a $\lambda/2$ cavity with a transparent dielectric mirror on the top and a AlAs/AlGaAs distributed Bragg reflector (DBR) at the bottom. The electric field profile due to positive potential at the transparent metal electrode, placed on the top of the QW, creates an electrically defined QD (as shown in Fig.1(a)). A single electron from the $n$-doped AlGaAs layer at the bottom is trapped in each QD [12]. The planar microcavity is resonant with the excitonic transition in the QW. As shown in Fig.1(b), in strong coupling regime, the normal states are the quasi-particles: lower polariton (LP) and upper polariton (UP), formed by the admixture of cavity photons and optically active heavy-hole (HH) excitons [13]. The splitting between the LP and UP branch depends on the strength of the coupling between the cavity and the exciton. The previous work by some of the present authors[14], described how a two-qubit controlled-$z$ gate operation can be carried out by a pump laser pulse focused in a spot covering two QDs. Before addressing the single qubit rotation along the $x$ (or $y$) axis, initialization and measurement techniques, we outline the principle of the control of electron spins with exciton-polaritons.

Consider an elliptically polarized pump laser incident at an angle $\theta_i$, focused over a single QD. The elliptical polarization is such that the refracted beam at the QW is circularly ($\sigma+$) polarized and has a circular cross-section of radius $= R$. The pump excites excitons with total angular momentum $J_{\theta_r} = 1$ (conservation of angular momentum) and $k_{||} = k_x = k\sin(\theta_r)$ (conservation of linear momentum along the plane of QW), where $\theta_r$ is the angle of refraction inside the QW. The subscript $\theta_r$ implies that the quantization axis is tilted from the growth direction (*i.e.*, $z$ axis) by $\theta_r$. Figure 1(c) shows the schematic of the QW and incident pump laser. As explained later, we will only consider heavy-hole (HH) excitons, which form polaritons with cavity photons of polarization $\sigma_+$. The optically excited excitons consist of electron spin $j_{\theta_{r,e}} = -1/2$ and hole spin $j_{\theta_{r,h}} = 3/2$. Our scheme virtually excites only LP with transverse momentum $k_x$, so the frequency of the laser $\omega_L$ is off-resonant (red detuned) from LP energy at $k_x$. We note that, the QW dark exciton resonance is far detuned from the pump laser and the electrostatic QD prevents the excitation of the trapped electron spin qubit to higher dipole moment exciton or trion states. Under these approximations, the spin dependent Coulomb exchange interaction of the spin qubit and the electronic part of the polariton, is represented by the Hamiltonian [8, 14]:

$$H_0 = \delta_p p^\dagger p - V r_{k_x}^2 s_{\theta_r} p^\dagger p + \Omega(t)(p^\dagger + p), \quad (1)$$

where $\delta_p$ is the polariton detuning from the laser, $p^\dagger(p)$ is the polariton creation(annihilation) operator, $s_{\theta_r}$ is the projection of the spin operators, for the localized electron onto the $\theta_r$ direction, $r_0$ is the Hopfield factor for the polaritons ($r_0 = 1/\sqrt{2}$ when the cavity is resonant with QW exciton energy) and $\Omega(t)$ is the external optical pumping rate. $V$ is the exchange coupling constant between the localized electron and the electronic part of the polariton, given as

$$V = \int d\mathbf{r_e} d\mathbf{r_h} d\mathbf{r_l} \frac{\psi(\mathbf{r_e}, \mathbf{r_h})\varphi(\mathbf{r_l}) e^2 \psi(\mathbf{r_l}, \mathbf{r_h})\varphi(\mathbf{r_e})}{4\pi\epsilon(|\mathbf{r_e} - \mathbf{r_l}|)}, \quad (2)$$

where $\epsilon$ is the dielectric constant of GaAs, $\mathbf{r_e}, \mathbf{r_h}$ are the position vectors of the electron and hole in the excitonic part of the polariton, $\mathbf{r_l}$ represents that of the localized electron, $\psi, \varphi$ represent the wavefunctions of the excitonic component of the polariton and localized electron. The Coulomb exchange interaction with the hole is very small and hence neglected [8, 15].

The effect of the $H_0$ in Eqn.1 is to rotate the electron spin about the axis defined by the angular momentum quantization axis of the polariton. Consequently, the rotation of the electron spin along the $x$ axis requires the excitation of the polaritons with the angular momentum oriented parallel to the QW plane. Unfortunately, such an orientation is prohibited because of the quantum confinement along the growth direction of the QW. The exciton is made up of: conduction band electron, with orbital angular momentum $\mathbf{L}_e = 0$ and spin angular momentum $\mathbf{S}_e = -1/2$; valence band hole, with $\mathbf{L_h} = 1$ and $\mathbf{S_h} = 1/2$. The quantization of $\mathbf{L}$ is along the linear momentum $\mathbf{k}$. However, in a QW with width $w$, $k_z = 2\pi/w >> k_{\|}$, which results in the quantization axis to be almost parallel to the growth direction [16]. The quantization direction can be evaluated by diagonalizing

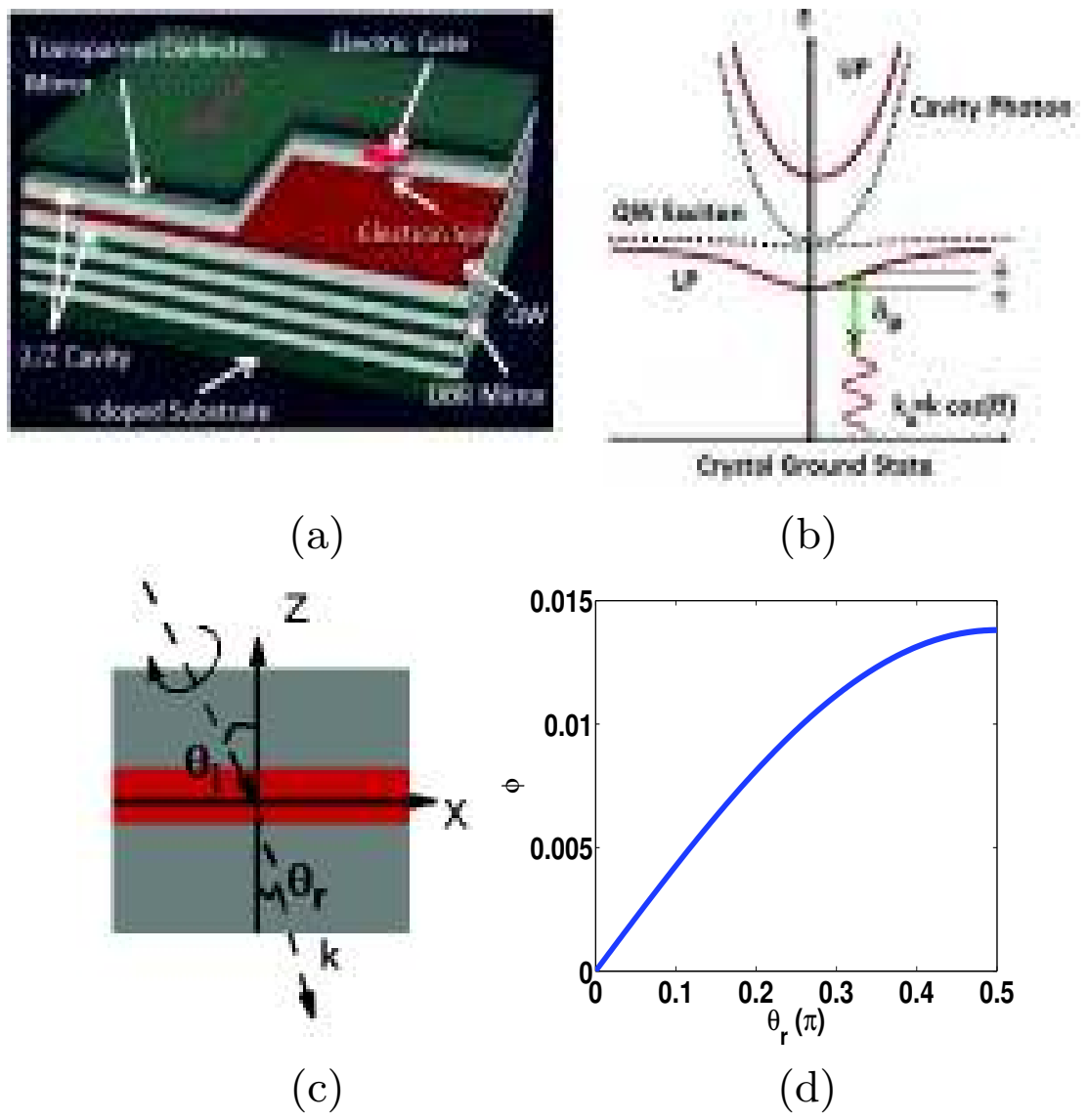

FIG. 1. (a) Illustration of the system consisting of QW placed in $\lambda/2$ DBR cavity. Electrons are trapped by the electric filed under the metal gates. Polaritons are formed in the region where the laser is incident. (b) Representation of the exciton polariton energy dispersion. The inset shows the spin dependent energy of the LP. (c) Schematic of the laser incident on the QW. (d) The variation of quantization direction $\phi$ with incidence angle $\theta_r$.

the Luttinger Hamiltonian of the valence band hole:

$$H_L = \begin{pmatrix} P+Q & L & M & 0 \\ L^* & P-Q & 0 & M \\ M^* & 0 & P-Q & -L \\ 0 & M^* & -L^* & P+Q \end{pmatrix}$$

where,

$$P = -\frac{\hbar^2}{2m_0}\gamma_1 k^2, \qquad Q = \frac{\hbar^2}{2m_0}\gamma_2(2k_z^2 - k_x^2),$$
$$L = \frac{\hbar^2}{2m_0} 2\sqrt{3}\gamma_3 k_x k_z \quad \& \quad M = \frac{\hbar^2}{2m_0}\sqrt{3}\frac{\gamma_2 + \gamma_3}{2} k_x^2.$$

Here $m_0$ is the mass of an electron and $\gamma_i$, $(i = 1, 2, 3)$ are material specific constants [16]. For the excitation wavelength of $\lambda \approx 786$ nm and $w = 10$ nm, the maximum $k_{\|} = 2\pi/\lambda \approx k_z/100$. In this limit, the light hole (LH) excitations are negligible, owing to the large HH-LH splitting. Figure 1(d) presents the dependence of the direction of quantization (represented by the angle $\phi$ it makes with the $z$ axis) on $\theta_r$. It is clear that even when the pump laser is directed along the plane of the QW, the excited polaritons have their angular momentum directed close to the $z$ axis ($\phi \approx 0.014$). In order to overcome this obstacle inherent to our device structure, we propose to use two co-planar ($xz$ plane) pump pulses, one incident from the left and the other from the right, having exactly the same intensity, spectral and temporal profile, but with opposite polarizations. The $\sigma_\pm$ polarized beams, incident at an angle of $\pm\theta_i$, excites a coherent polariton population with $J_{\pm\theta_r} = 1$ and $k_\parallel = \pm k_x$. The angular momentum quantization axis for the polaritons with linear momentum $\pm k_x$ makes an angle $\pm\phi$ with the $z$ axis, which implies that polaritons with $J_{\pm\phi} = \pm 1, 0$ are excited. However, the $J_{\pm\phi} = 0$ polaritons originate from LHs and as mentioned previously, do not contribute to the interaction. In the absence of localized electron spin, the polariton fields will superimpose, resulting in an effective field polarization along the x axis, with no net angular momentum.

In the presence of electron spin qubit this process is more complicated. Nonetheless, the effective axis of rotation can be steered parallel to the $x$ axis by optimizing the polariton-laser detuning $\delta_p$ and the incidence angle $\theta_i$ or equivalently $\theta_r$. The Hamiltonian for the interaction can now be written as:

$$\begin{aligned} H_0 = & \ \delta_p \sum_{m,n} p^\dagger_{m,n} p_{m,n} - 2V r^2_{k_x} \cos(\phi) s_{ze} \sum_{m,n} m p^\dagger_{m,n} p_{m,n} \\ & + 2V r^2_{k_x} \sin(\phi) s_{xe} \sum_{m,n} mn p^\dagger_{m,n} p_{m,n} \\ & + \Omega(t) \sum_{m,n} \frac{1 + mn\cos(\phi + \theta_r)}{2} (p_{m,n} + p^\dagger_{m,n}) \end{aligned} \tag{3}$$

where, $p^\dagger_{m,n}(p_{m,n})$ is the creation(annihilation) operator for the polariton. The second term in the subscript defines the polariton's in plane linear momentum $k_\parallel = nk_x (n = \pm 1)$ and the first term $(m = \pm 1)$ defines the angular momentum projection $J_{n\phi} = m (m = \pm 1)$. The second term containing the electron spin operator $s_{ze}$ rotates the qubit about the $z$ axis, whereas the third term with $s_{xe}$ tries to rotate it about the $x$ axis. From now onwards the spin of the trapped electron will be quantized along the $z$. From the last two pump terms it is clear that rate at which the polaritons $p_{-1,1}(p_{1,1})$ are being injected in the QW is the same as that for polaritons $p_{1,-1}(p_{-1,-1})$. Thus, if the evolution of the polaritons in the system is slow enough, so that population of $p_{-1,1}(p_{1,1})$ polaritons remains the same as that of $p_{1,-1}(p_{-1,-1})$ polaritons, the second term will vanish and the rotation will be purely about the $x$ axis. In the presence of decoherence introduced by the finite lifetime of polaritons in the cavity, the system is a mixture of states $|1\rangle = |1/2, \alpha_{1,1}, \alpha_{-1,1}, \alpha_{1,-1}, \alpha_{-1,-1}\rangle$ and $|2\rangle = |-1/2, \beta_{1,1}, \beta_{-1,1}, \beta_{1,-1}, \beta_{-1,-1}\rangle$ and is best represented by the density operator $\rho$. $|1\rangle(|2\rangle)$ represents the state with the localized electron spin with $s_z = 1/2(-1/2)$ and the polaritons in the coherent state $\alpha_{\pm1,\pm1}(\beta_{\pm1,\pm1})$, such that $p_{m,n}|\alpha_{m,n}\rangle = \alpha_{m,n}|\alpha_{m,n}\rangle(p_{m,n}|\beta_{m,n}\rangle = \beta_{m,n}|\beta_{m,n}\rangle)$, $m, n = \pm 1$. The master equation governing the evolution of the density matrix is given in the Lindblad form:

$$\frac{d\rho}{dt} = -i[H_0, \rho] + \gamma \sum_{m,n} p_{m,n} \rho p^\dagger_{m,n} - \frac{\gamma}{2} \sum_{m,n} \{p^\dagger_{m,n} p_{m,n}, \rho\}. \tag{4}$$

Here, $\gamma$ represents the rate at which polaritons are lost from the cavity and depends on the reflectivity of the cavity mirrors (r) and the radius (R) of the laser spot. Since the Coulomb exchange interaction also depends on $R$, there is unique relationship between $V$ and $\gamma$ [14, 17]. If one tries to increase $V$ by tighter focusing of the laser spot, $\gamma$ will increase. Any attempt to decrease the leakage from cavity by increasing $R$ will lead to a decrease in $V$. The optimal regime that we choose is when $R = 6$ $\mu$m, resulting in $V = 0.2$ $\mu$eV and $\gamma = 0.3$ meV, which can be easily reached in a cavity with nominal $Q \approx 3000$.

If the spin qubit was initialized to $|-1/2\rangle$, then after a rotation of $\pi$ about the $x$ axis, its final state will be $|1/2\rangle$. The fidelity $F$ of this operation is:

$$F = Tr[\sqrt{\rho_a} \rho_e \sqrt{\rho_a}] = \langle 1|\rho|1\rangle, \tag{5}$$

where, $\rho_a$ is the desired final density matrix and $\rho_e$ is the real density matrix obtained by considering errors. $F$ will be 1 when the localized electron spin undergoes a perfect rotation by $\pi$ around the $x$ axis. As explained previously, the rotation axis depends on the relative population of the polaritons $p_{m,n}$. So in general, the localized spin rotates about an axis in the $xz$ plane. In addition, the loss of cavity polaritons causes decoherence and hence $F < 1$. The variation of $F$ with $\theta_r$ is shown in Fig.2(a). This dependence can be understood from Fig.**??**, which shows that $\phi$ increases with $\theta_r$ and hence from the third term in Eq. (3), the strength of rotation about the $x$ axis, and consequently $F$, increases. Note that $r$ decreases outside the stop-band of the DBR mirror. In addition, $r_k$ increases as $k_x$ (or $\theta_r$) increases, making the LP more exciton like. So we confine our analysis for to small $theta_r$s only. The fidelity will be high, if the interaction $(V)$ of polaritons with the spin qubits is slow enough, so that qubits are able to *see* the superimposed field of the polaritons, with *effective* polarization axis along the $x$ direction. Nonetheless, if $V$ decreases too much then there will be no interaction at all and hence no rotation. For a fixed $V$, the rate of evolution of the polaritons can be lowered by increasing the detuning $\delta_p$ and so $F$ increases with $\delta_p$. This dependence of $F$ on $V$ and $\delta_p$ is depicted in Fig.2(b) and (c). By examining these plots we identify the practical parameter regime to achieve high $F$ as $\theta_r = 0.1$ rad ($\equiv \theta_i = pi/6$ rad, $r_{k_x} = 0.75$), $V = 0.25$ $\mu$eV, $\gamma = 0.3$ eV (corresponding to $r = 0.999$) and $\delta_p = 6$ meV. For this set of optimal parameters, the dependence of $F$ on the pump pulse energy $(\Omega_0)$ and duration $(\tau)$ is displayed in Fig.2(d). We have assumed a Gaussian pump profile so that $\Omega(t) = \Omega_0 e^{-t^2/\tau^2}$. $\langle 1|\rho|1\rangle$ and $\langle 3|\rho|3\rangle$ depend on the area of the phase space trajectory of the coherent states $\alpha_{m,n}, \beta_{m,n}$. The width of the path depends on $\tau$ and its length depends on $\Omega_0$. So, to have the path corresponding to a fixed $F$, if $\Omega_0$ increases then $\tau$ must decrease to keep the area constant. This explains the fidelity pattern in Fig.2(d). We are able to obtain $F = 99.94\%$ with $\Omega_0 = 600$ meV and $\tau = 312$ ps, which corresponds to a moderate pump power of $P = \Omega(t)^2 \omega \tau_{photon}/(\pi R^2 \hbar) = 7.9$ mW/$\mu$m$^2$ [14], where $\tau_{photon}$ is the photon lifetime in the cavity. The single qubit gate time for the $\pi$ rotation is $\approx 4\tau \approx 1.2$ ns. Since the cavity has a finite lifetime, the leaked photons can disclose the information about the state of the qubit, where $\Re$ and $\Im$ are the real and imaginary parts. We note that the vacuum field fluctuation leads to a uncertainty in the coherent state, so that, $\Delta(\Im[\alpha_{m,n}, \beta_{m,n}])^2 = \Delta(\Re[\alpha_{m,n}, \beta_{m,n}])^2 = 1/4$. In the parameter regime listed above, the distance between the phase space trajectories of the leaked photons entangled with the localized electron having spin $s_z = 1/2$ and $s_z = -1/2$ is much less than the quantum noise [17]. Thus, a quantum erasure is established and the information is protected from an eavesdropper. With the parameters used for single qubit rotation, a two-qubit controlled-$z$ operation, as described in [14], can be achieved with fidelity 99.99% in 1.4 ns, with a pump power of $P = 2.6$ mW/$\mu$m$^2$. As it was mentioned before, due to the inability to create LPs with angular momentum quantized along the plane of the QW, $F$ of the single qubit rotation is lower than that of the two-qubit operation in the practical parameter regime.

Apart from being able to perform fast universal logic operations on a scalable system, this scheme offers the advantage of a projective measurement. If an off resonant laser is focussed along the $z$ axis, over a single qubit, then the reflected coherent light will develop a phase depending on the state of the qubit. The laser has low power and is unable to cause any rotation of the qubit. By performing a Homodyne measurement of the small, spin dependent phase shift in the reflected light, one can determine the state of the qubit. This measurement is projective and can be used to initialize the qubit.

An electron spin trapped in GaAS QW is subjected to hyperfine interaction with the nucleus and thus has a small $T_2^* \approx 10$ ns. By using spin echo technique [18, 19], $T_2^* \approx 3$ $\mu$s has been achieved. The principle of spin echo can be extended to our proposal. Two lasers in the $xz$ plane rotate the spin by $\pi/2$ about the $x$ axis in time $t_{\pi/2}$. Next, these two lasers are switched off, while the electron spin is allowed to precesses freely in the $xy$ plane for a time $T$. After time $T$, two other lasers in the $yz$ plane cause a spin flip. Finally, the lasers in $xz$ plane are again activated to rotate the spin by $\pi/2$ in time $t_{\pi/2}$. In our scheme, the single qubit and two-qubit operations have an error rate of $6 \times 10^-4$ and $1 \times 10^-4$ respectively. The low error rates and long-ranged interaction make it a favorable choice in fault tolerant quantum computing architecture [20, 21].

In conclusion, we described a complete scheme for all optical, initialization, single qubit and two-qubit operations, and measurement, using Coulomb exchange interaction between trapped electron spin qubits and cavity polaritons. The single qubit rotation is about 100 times slower than other proposals, however the main advantage is that, in the same set-up, high speed, high fidelity two-qubit operations can be achieved using practical pump powers and cavity $Q(\approx 3000)$. With a large separation ($\approx 6$ $\mu$m) between the qubits, the two-qubit gate operation time is $10^6$ faster than the current proposals [22, 23]. We believe that this is a promising scheme for a practical realization of universal quantum computation.

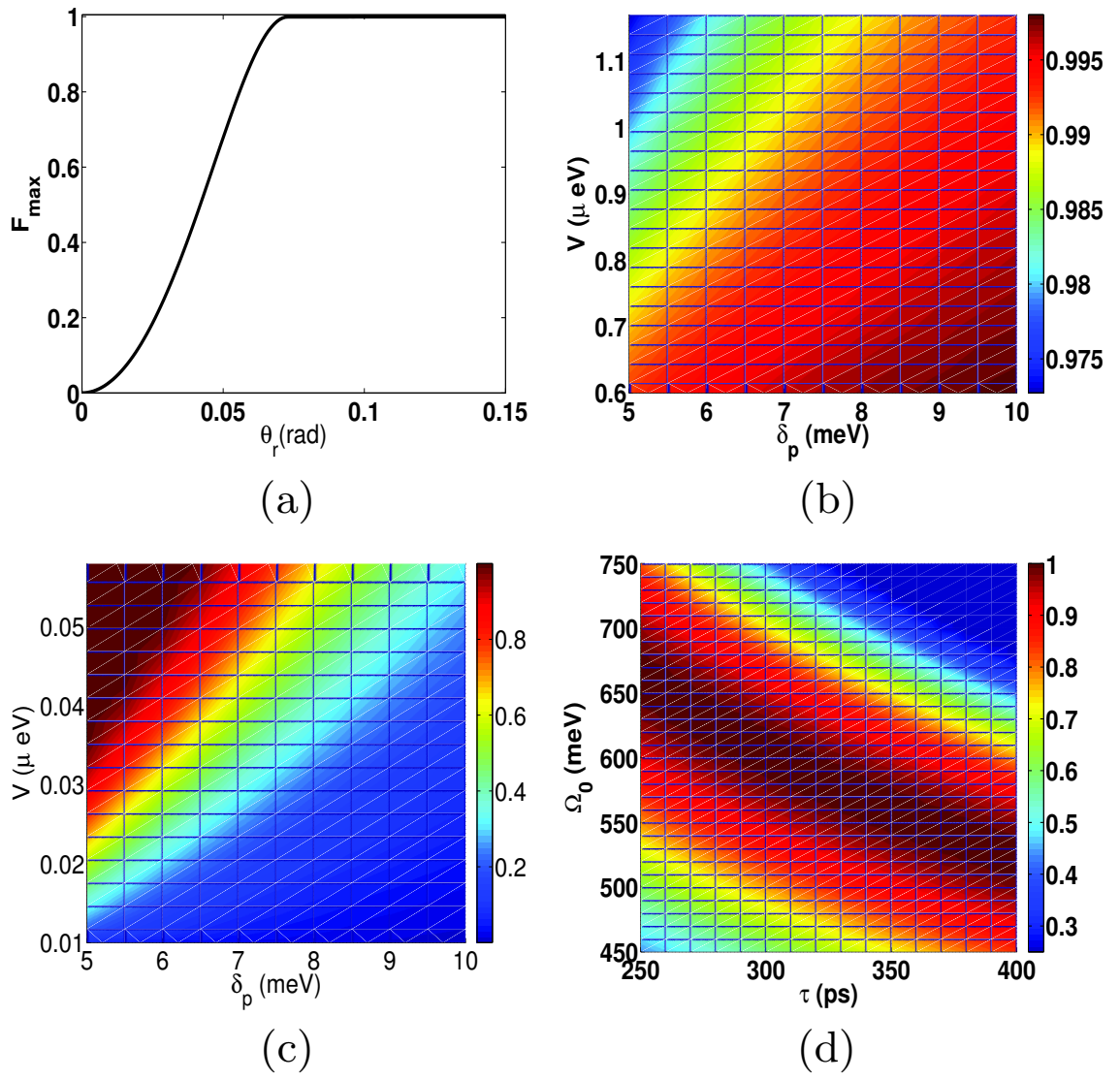

FIG. 2. Maximum $F$ achievable for different (a) angles $\theta_r$, when $\delta_p = 5$ meV, $V = 0.2$ $\mu$eV and $\gamma = 0.3$ meV. (b),(c) for $V$ and $\delta$ for $\theta_r = \pi/6$ , (d) $\Omega_0$ and $\tau$, for optimal parameters $\theta_r = \pi/6$, $V = 0.2$ $\mu$eV, $\gamma = 0.3$ eV and $\delta_p = 5$ meV.

## ACKNOWLEDGMENTS

This work has been supported by the Japan Society for the Promotion of Science(JSPS) through its Funding Program for World-Leading Innovative R&D on Science and Technology (FIRST Program)" and NICT. The authors appreciate Dr. G. F. Quinteiro, Dr. T. D. Ladd and Dr. Y. C. N. Na for insightful discussions.